\documentclass[aps,prd,twocolumn,nofootinbib,,superscriptaddress,showkeys]{revtex4}
\usepackage{amssymb,amsmath,graphicx,enumerate}
\usepackage{subfigure}
\usepackage{dcolumn,booktabs}

\usepackage[dvipdfm,colorlinks=true,citecolor=blue,pdfstartview=FitH]{hyperref}

\usepackage{color,framed} 
\definecolor{shadecolor}{rgb}{1,0,0} 

\def\bea{\begin{equation}}
\def\eea{\end{equation}}
\newcommand{\thbs}{triply heavy bottom-charm baryons}
\newcommand{\thb}{triply heavy bottom-charm baryon}
\newcommand{\thbf}{$\Omega_{ccb}$ and $\Omega_{cbb}$}
\newcommand{\rt}{Regge trajectory}
\newcommand{\rts}{Regge trajectories}

\newcommand{\trs}{trajectories}
\newcommand{\bfr}{{\bf r}}
\newcommand{\bfp}{{\bf p}}
\newcommand{\bfpa}{{|\bf p|}}
\newcommand{\gev}{{\rm GeV}}

\newcommand{\qp}{{q^{\prime}}}
\newcommand{\qpp}{{q^{\prime\prime}}}

\newcommand{\qqs}{{[qq^{\prime}]}}
\newcommand{\qqb}{\{qq^{\prime}\}}
\newcommand{\cltb}{$\bar{3}_c$}
\newcommand{\cltba}{\bar{3}_c}

\newcommand{\dqs}{$(qq')$}
\newcommand{\bqm}{\,{\pmb{?}}}

\begin{document}
\title{Regge trajectories for the triply heavy bottom-charm baryons in the diquark picture}
\author{Jia-Qi Xie}
\email{1462718751@qq.com}
\affiliation{School of Physics and Information Engineering, Shanxi Normal University, Taiyuan 030031, China}
\author{He Song}
\email{songhe\_22@163.com}
\affiliation{School of Physics and Information Engineering, Shanxi Normal University, Taiyuan 030031, China}
\author{Jiao-Kai Chen}
\email{chenjk@sxnu.edu.cn, chenjkphy@outlook.com (corresponding author)}
\affiliation{School of Physics and Information Engineering, Shanxi Normal University, Taiyuan 030031, China}

\begin{abstract}
We present the explicit form of the Regge trajectory relations for the triply heavy bottom-charm baryons, which can be applied to investigate both the $\lambda$-mode excited states and the $\rho$-mode excited states. We estimate the masses of the $\lambda$-excited states and the $\rho$-excited states. The results are in agreement with other theoretical predictions. Both the $\lambda$-trajectories and the $\rho$-trajectories are discussed.
Moreover, the behaviors of the $\lambda$- and $\rho$-trajectories for various baryons are discussed.
It is shown that both the $\lambda$-trajectories and the $\rho$-trajectories for baryons are concave downwards in the $(M^2,\,x)$ plane. The Regge trajectories for the light baryons are approximately linear and become concave as the masses of the light constituents are considered.
\end{abstract}

\maketitle


\section{Introduction}
In the diquark picture, the {\thbs}, $\Omega_{ccb}$ and $\Omega_{cbb}$, are composed of one doubly diquark ($(cc)$ or $(bb)$) and one heavy quark ($b$ or $c$). The {\thbs} have been studied by a vast variety of approaches including the lattice QCD \cite{Mathur:2018epb,Brown:2014ena}, the renormalization group procedure for effective particles \cite{Serafin:2018aih}, the relativistic quark model based on the quark-diquark picture in the quasipotential approach \cite{Faustov:2021qqf}, the non-relativistic quark model \cite{Silvestre-Brac:1996myf,Roberts:2007ni,deArenaza:2024dhe}, the constituent quark model \cite{Yang:2019lsg}, the bag model \cite{Hasenfratz:1980ka}, the sum rules \cite{Zhang:2009re,Najjar:2024deh,Wang:2011ae}, the relativistic three-quark model \cite{Martynenko:2007je}, the variational method \cite{Jia:2006gw}, and baryon Faddeev equations \cite{Yin:2019bxe}.

The {\thbs} are also discussed by the {\rts} \footnote{A {\rt} of hadrons is assumed to be written as $M=m_R+\beta_x(x+c_0)^{\nu}$ $(x=l,\,n_r)$, where $M$ is mass of the bound state, $l$ is the the orbital angular momentum, and $n_r$ is the radial quantum number. $m_R$ and $\beta_x$ are parameters. For simplicity, the plots in the $(M,\,x)$ plane \cite{Chen:2022flh,Chen:2023cws}, in the $(M,\,(x-c_0)^{\nu})$ plane \cite{Burns:2010qq}, in the $(M^2,\,x)$ plane \cite{Chen:2018nnr,Chen:2021kfw}, in the $((M-m_R)^2,\,x)$ plane \cite{Chen:2023djq,Chen:2023web} or in the $((M-m_R)^{1/{\nu}},\,x)$ plane \cite{Xie:2024dfe} are all called the Chew-Frautschi plots. The {\rts} can be plotted in these different planes. } \cite{Ishida:1994pf,Chen:2023djq,Oudichhya:2023pkg}.
In preceding works, the {\rts} for the {\thbs} are mostly the $\lambda$-{\trs}, with few studies addressing the $\rho$-{\trs} \cite{Ishida:1994pf}. In this work, we investigate both the $\lambda$-{\trs} and the $\rho$-{\trs} for the {\thbs}.
The {\rts} take different forms in different energy regions \cite{Chen:2022flh,Chen:2021kfw}. In the case of the $\lambda$-mode, the {\thbs} are the heavy-heavy systems because both the diquark ($(bb)$ or $(cc)$) and the quark ($b$ or $c$) are heavy. In the case of the $\rho$-mode, the {\thbs} also show the heavy-heavy properties because the diquarks $(bb)$ and $(cc)$ are the heavy-heavy systems.
By employing the diquark {\rt} relation \cite{Feng:2023txx}, we present the {\rt} relations (\ref{t2q}) and (\ref{combrt}) for the {\thbs}. The masses of the $\lambda$-excited states and the $\rho$-excited states are estimated, and both the $\lambda$-{\trs} and the $\rho$-{\trs} are discussed.

The paper is organized as follows: In Sec. \ref{sec:rgr}, the {\rt} relations are obtained from the spinless Salpeter equation. In Sec. \ref{sec:rtdiquark}, we investigate the {\rts} for the {\thbs} and estimate the masses of the excited states. The conclusions are presented in Sec. \ref{sec:conc}.

\section{{\rt} behaviors for various baryons}\label{sec:rgr}

The {\rts} for different systems take different forms \cite{Chen:2022flh,Chen:2021kfw}. In this section, we use the diquark {\rts} \cite{Chen:2023cws,Feng:2023txx,Chen:2023ngj} to discuss the behaviors of the baryon {\rts}. We show that the behaviors of the {\rts} for the triply heavy baryons are different from the behaviors of the {\rts} for other types of baryons.
The behaviors of the {\rts} for various tetraquarks are referred to Ref. \cite{Xie:2024dfe}.

\subsection{Preliminary}\label{subsec:prelim}

\begin{figure}[!phtb]
\centering
\includegraphics[width=0.25\textheight]{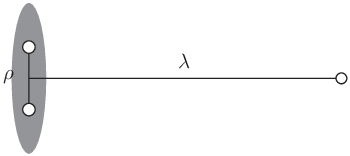}
\caption{Schematic diagram of the baryons in the diquark-quark picture.}\label{fig:tr}
\end{figure}

In the diquark picture, baryons consist of one quark in color $3_c$ and one diquark in color $\bar{3}_c$, see Fig. \ref{fig:tr}. $\rho$ separates the quarks in the diquark, and $\lambda$ separates the quark and the diquark. There exist two excited modes: the $\rho$-mode involves the radial and orbital excitation in the diquark, and the $\lambda-$mode involves the radial or orbital excitation between the quark and diquark. Consequently, there exist two series of {\rts}: one series of $\rho$-{\trs} and one series of $\lambda$-{\trs}.

\begin{table}[!phtb]
\caption{The completely antisymmetric states for the diquarks in {\cltb} \cite{Feng:2023txx}. $j_d$ is the spin of the diquark {\dqs}, $s_d$ denotes the total spin of two quarks, $l$ represents the orbital angular momentum. $n=n_r+1$, $n_r$ is the radial quantum number, $n_r=0,1,2,\cdots$. }  \label{tab:dqstates}
\centering
\begin{tabular*}{0.49\textwidth}{@{\extracolsep{\fill}}ccccc@{}}
\hline\hline
 Spin of diquark & Parity  &  Wave state  &  Configuration    \\
( $j_d$ )          & $(j_d^P)$ & $(n^{2s_d+1}l_{j_d})$  \\
\hline
$j_d=0$              & $0^+$   & $n^1s_0$         & $[qq']^{{\cltba}}_{n^1s_0}$ \\
                 & $0^-$   & $n^3p_0$         & $[qq']^{{\cltba}}_{n^3p_0}$       \\
$j_d=1$              & $1^+$   & $n^3s_1$, $n^3d_1$   & $\{qq'\}^{{\cltba}}_{n^3s_1}$,\;    $\{qq'\}^{{\cltba}}_{n^3d_1}$\\
                 & $1^-$   & $n^1p_1$, $n^3p_1$   &
$\{qq'\}^{{\cltba}}_{n^1p_1}$,\; $[qq']^{{\cltba}}_{n^3p_1}$ \\
$j_d=2$              & $2^+$   & $n^1d_2$, $n^3d_2$         &  $[qq']^{{\cltba}}_{n^1d_2}$,\; $\{qq'\}^{{\cltba}}_{n^3d_2}$\\
                 & $2^-$   & $n^3p_2$, $n^3f_2$       &
 $[qq']^{{\cltba}}_{n^3p_2}$,\; $[qq']^{{\cltba}}_{n^3f_2}$       \\
$\cdots$         & $\cdots$ & $\cdots$               & $\cdots$  \\
\hline\hline
\end{tabular*}
\end{table}

In the diquark picture, the state of a baryon is denoted as
\bea\label{tetnot}
\left((qq')_{n^{2s_d+1}l_{j_d}}
{\qpp}\right)_{N^{2j+1}L_J},
\eea
where the superscripts {\cltb}, $3_c$ and $1_c$ are omitted. The diquark $(qq')$ is $\{qq'\}$ or $[qq']$. $\{qq'\}$ and $[qq']$ indicate the permutation symmetric and antisymmetric flavor wave functions, respectively. The completely antisymmetric states for the diquarks in {\cltb} are listed in Table \ref{tab:dqstates}. $N=N_{r}+1$, where $N_{r}=0,\,1,\,\cdots$. $n=n_{r}+1$, where $n_{r}=0,\,1,\,\cdots$. $N_r$ and $n_{r}$ are the radial quantum numbers of the baryon and diquark, respectively.
$\vec{J}=\vec{L}+\vec{j}$, $\vec{j}=\vec{j}_d+\vec{s}_{\qpp}$, $\vec{j}_d=\vec{s}_d+\vec{l}$.
$\vec{J}$, $\vec{j}_d$ and $\vec{s}_{\qpp}$ are the spins of baryon, diquark and quark $q^{\prime\prime}$, respectively. $\vec{j}$ is the summed spin of diquark and quark in the baryon. $L$ and $l$ are the orbital quantum numbers of baryon and diquark, respectively. $\vec{s}_{d}$ is the summed spin of quarks in the diquark.

\subsection{Spinless Salpeter equation}
The spinless Salpeter equation \cite{Godfrey:1985xj,Ferretti:2019zyh,Bedolla:2019zwg,Durand:1981my,Durand:1983bg,Lichtenberg:1982jp,Jacobs:1986gv} reads as
\begin{eqnarray}\label{qsse}
M\Psi_{d,b}({\bfr})=\left(\omega_1+\omega_2\right)\Psi_{d,b}({\bfr})+V_{d,b}\Psi_{d,b}({\bfr}),
\end{eqnarray}
where $M$ is the bound state mass (diquark or baryon). $\Psi_{d,b}({\bfr})$ are the diquark wave function and the baryon wave function, respectively. $V_{d,b}$ denotes the diquark potential and the baryon potential, respectively (see Eq. (\ref{potv})). $\omega_1$ is the relativistic energy of constituent $1$ (quark or diquark), and $\omega_2$ is of constituent $2$ (quark),
\bea\label{omega}
\omega_i=\sqrt{m_i^2+{\bf p}^2}=\sqrt{m_i^2-\Delta}\;\; (i=1,2).
\eea
$m_1$ and $m_2$ are the effective masses of constituent $1$ and $2$, respectively.

When using diquark in multiquark systems, the interactions between quark and quark, diquark and quark, and diquark and diquark are needed. In Ref. \cite{Faustov:2021hjs}, these interactions are constructed with the help of the off-mass-shell scattering amplitude, which is projected onto the positive energy states.
The interactions can also be established by expanding the interactions of the quark-antiquark system to the quark-quark system, and then to the diquark-antidiquark systems or the diquark-quark systems \cite{Lundhammar:2020xvw}.
Furthermore, the effect of the finite size of diquark is treated differently. In Refs. \cite{Faustov:2021hjs}, the size of diquark is taken into account through corresponding form factors. At times, diquark is taken as being pointlike \cite{Ferretti:2019zyh,Lundhammar:2020xvw}; we use this approximation in the present work.

Following Refs. \cite{Ferretti:2019zyh,Bedolla:2019zwg,Ferretti:2011zz,Eichten:1974af,
Chen:2023web,Chen:2023djq}, we employ the potential
\begin{align}\label{potv}
V_{d,b}&=-\frac{3}{4}\left[V_c+{\sigma}r+C\right]
\left({\bf{F}_i}\cdot{\bf{F}_j}\right)_{d,b},
\end{align}
where $V_c\propto{1/r}$ is a color Coulomb potential or a Coulomb-like potential due to one-gluon-exchange. $\sigma$ is the string tension. $C$ is a fundamental parameter \cite{Gromes:1981cb,Lucha:1991vn}. The part in the bracket is the Cornell potential \cite{Eichten:1974af}. ${\bf{F}_i}\cdot{\bf{F}_j}$ is the color-Casimir,
\bea\label{mrcc}
\langle{(\bf{F}_i}\cdot{\bf{F}_j})_{d}\rangle=-\frac{2}{3},\quad
\langle{(\bf{F}_i}\cdot{\bf{F}_j})_{b}\rangle=-\frac{4}{3}.
\eea

\subsection{{\rt} relations for different systems}
In this subsection, we present the {\rt} relations for the heavy-heavy systems, the heavy-light systems, and the light systems.

For the heavy-heavy systems, $m_{1},m_2{\gg}{\bfpa}$, Eq. (\ref{qsse}) reduces to
\begin{eqnarray}\label{qssenrr}
M\Psi_{d,t}({\bfr})&=&\left[(m_1+m_2)+\frac{{\bfp}^2}{2\mu}\right]\Psi_{d,t}({\bfr})\nonumber\\
&&+V_{d,t}\Psi_{d,t}({\bfr}),
\end{eqnarray}
where
\bea\label{rdmu}
\mu=m_1m_2/(m_1+m_2).
\eea
By employing the Bohr-Sommerfeld quantization approach \cite{Brau:2000st} and using Eqs. (\ref{potv}) and (\ref{qssenrr}), we obtain the parameterized relation \cite{Chen:2022flh,Chen:2021kfw}
\bea\label{massform}
M=m_R+\beta_x(x+c_{0x})^{2/3}\;(x=l,\,n_r,\,L,\,N_r)
\eea
with
\bea\label{rtft}
m_R=m_1+m_2+C',
\eea
where
\bea\label{cprime}
C'=\left\{\begin{array}{cc}
C/2, & \text{diquarks}, \\
C, & \text{baryons}.
\end{array}\right.
\eea
\bea\label{sigma}
\sigma'=\left\{\begin{array}{cc}
\sigma/2, & \text{diquarks}, \\
\sigma, & \text{baryons}.
\end{array}\right.
\eea
$\beta_x$ reads
\bea\label{parabm}
\beta_x=c_{fx}c_xc_c.
\eea
$c_c$ and $c_x$ are
\bea
c_c=\left(\frac{\sigma'^2}{\mu}\right)^{1/3},\; c_l=\frac{3}{2},\; c_{n_r}=\frac{\left(3\pi\right)^{2/3}}{2}.
\eea
Both $c_{fl}$ and $c_{fn_r}$ are equal theoretically to one and are fitted in practice.
In Eq. (\ref{massform}), $m_1$, $m_2$, $c_x$ and $\sigma$ are universal. $c_{0x}$ are determined by fit. $c_{fl}$ and $c_{fn_r}$ are theoretically equal to one but are fitted in practice.

For the heavy-light systems ($m_1\to\infty$ and $m_2\to0$), Eq. (\ref{qsse}) simplifies to
\begin{eqnarray}\label{qssenr}
M\Psi_{d,t}({\bfr})=\left[m_1+{\bfpa}+V_{d,t}\right]\Psi_{d,t}({\bfr}).
\end{eqnarray}
By employing the Bohr-Sommerfeld quantization approach \cite{Brau:2000st}, the parameterized formula can be obtained and is written as \cite{Chen:2022flh,Chen:2021kfw}
\bea\label{rtmeson}
M=m_R+\beta_x\sqrt{x+c_{0x}}\;(x=l,\,n_r,\,L,\,N_r).
\eea
The parameters in Eq. (\ref{rtmeson}) read as
\bea\label{massformhl}
c_{c}=\sqrt{\sigma'},\; c_l=2,\; c_{n_r}=\sqrt{2\pi}.
\eea
For the heavy-light systems, the common choice of $m_R$ is \cite{Selem:2006nd,Chen:2021kfw,Veseli:1996gy}
\bea\label{mrm1}
m_R=m_1.
\eea

Eqs. (\ref{rtmeson}) and (\ref{mrm1}) are obtained in the limit $m_1\to\infty$ and $m_2\to0$.
There are different ways to include the light constituent's mass \cite{Selem:2006nd,Nielsen:2018uyn,Sonnenschein:2018fph,
MartinContreras:2020cyg,Chen:2023cws,Chen:2023ngj,Chen:2022flh,Chen:2014nyo,
Afonin:2014nya,Sergeenko:1994ck}. Two modified formulas are proposed in Ref. \cite{Chen:2023cws}, which can universally describe both the heavy-light mesons and the heavy-light diquarks. One is Eq. (\ref{rtmeson}) together with Eq. (\ref{rtft}).
Another reads
\bea\label{mrtf}
M=m_R+\sqrt{\beta_x^2(x+c_{0x})+\kappa_{x}m^{3/2}_2(x+c_{0x})^{1/4}}
\eea
if $m_2{\ll}M$, where
\bea\label{mrfp}
m_R=m_1+C',\quad \kappa_x=\frac{4}{3}\sqrt{{\pi}\beta_x},
\eea
where $\beta_x$ is in (\ref{massformhl}).
Eqs. (\ref{rtmeson}) and (\ref{rtft}) are the extension of \cite{Afonin:2014nya}
\bea\label{afoninequ}
M=m_1+m_2+\sqrt{a(n_r+{\alpha}l+b)}
\eea
and the formula \cite{Chen:2022flh}
\bea\label{rmfnpb}
(M-m_1-m_2-C)^2=\alpha_x(x+c_0)^{\gamma}
\eea
while
(\ref{mrtf}) and (\ref{mrfp}) are based on the results in \cite{Selem:2006nd,Sonnenschein:2018fph}.
Although they give different behavior of $m_2$, Eqs. (\ref{rtmeson}), (\ref{rtft}) , (\ref{mrtf}) and (\ref{mrfp}) produce consistent results for $l,\,n_r<10$ and have the same behavior $M{\sim}x^{1/2}$ \cite{Chen:2023cws}.

For the light systems ($m_1,\,m_2\to0$), Eq. (\ref{qsse}) simplifies to
\begin{eqnarray}\label{qsseur}
M\Psi_{d,t}({\bfr})=\left[2{\bfpa}+V_{d,t}\right]\Psi_{d,t}({\bfr}).
\end{eqnarray}
By employing the Bohr-Sommerfeld quantization approach \cite{Brau:2000st}, the parameterized formula can be obtained, which is written as \cite{Chen:2022flh,Chen:2021kfw}
\bea\label{rtmesonur}
M=\beta_x\sqrt{x+c_{0x}}\;(x=l,\,n_r,\,L,\,N_r).
\eea
For the light systems, the parameters read as
\bea\label{massformhl}
c_{c}=\sqrt{\sigma'},\; c_l=2\sqrt{2},\; c_{n_r}=2\sqrt{\pi}.
\eea
According to the results in Refs.  \cite{Selem:2006nd,Sonnenschein:2018fph}, similar to Eq. (\ref{mrtf}) proposed in \cite{Chen:2023cws}, we suggest
\bea\label{lightm3}
M=C'+\sqrt{\beta_x^2(x+c_{0x})+\kappa_{x}\left(m_1^{3/2}+m_2^{3/2}\right)
(x+c_{0x})^{1/4}}.
\eea
Based on Eq. (\ref{afoninequ}) \cite{Afonin:2014nya}, we obtain a modified {\rt} relation for the light-light systems in Ref. \cite{Chen:2023ngj}
\bea\label{lighlight}
M=m_R+\beta_x\sqrt{x+c_{0x}},
\eea
where
\bea
c_c=\sqrt{\sigma'},\; c_l= 2\sqrt{2},\; c_{n_r}=2\sqrt{\pi}.
\eea
$m_R$ is in Eq. (\ref{rtft}).

\begin{table}[!phtb]
\caption{The coefficients for the heavy-heavy systems (HHS), the heavy-light systems (HLS), and the light-light systems (LLS).}  \label{tab:eparam}
\centering
\begin{tabular*}{0.47\textwidth}{@{\extracolsep{\fill}}cccc@{}}
\hline\hline
                   & HHS &  HLS & LLS  \\
\hline
$\nu$    & $2/3$ & $1/2$  & $1/2$  \\
$c_c$    & $\left({\sigma'^2}/{\mu}\right)^{1/3}$    & $\sqrt{\sigma'}$ & $\sqrt{\sigma'}$   \\
$c_{l,\,L}$    & $3/2$ & $2$ & $2\sqrt{2}$  \\
$c_{n_r,\,N_r}$ & ${\left(3\pi\right)^{2/3}}/{2}$      & $\sqrt{2\pi}$   & $2\sqrt{\pi}$ \\
\hline
\hline
\end{tabular*}
\end{table}

When Eqs. (\ref{massform}) and (\ref{rtft}) are applied to discuss the heavy-heavy systems, Eqs. (\ref{rtmeson}) and (\ref{rtft}) are employed to discuss the heavy-light systems, and Eqs. (\ref{lighlight}) and (\ref{rtft}) are used to discuss the light systems, summarising Eqs. (\ref{massform}), (\ref{rtft}), (\ref{rtmeson}) and (\ref{lighlight}), we have a general form of the {\rts} \cite{Chen:2022flh,Xie:2024dfe}
\begin{align}\label{massfinal}
M=&m_R+\beta_x(x+c_{0x})^{\nu}\,\,(x=l,\,n_r,\,L,\,N_r),\nonumber\\
m_R=&m_1+m_2+C',\quad \beta_x=c_{fx}c_xc_{c},
\end{align}
where ${\nu}$, $c_x$ and $c_{c}$ are listed in Table \ref{tab:eparam}. $c_{fx}$ are theoretically equal to one and are fitted in practice. $c_{0x}$ vary with different {\rts}. Eq. (\ref{massfinal}) can be employed to discuss various systems including the heavy-heavy systems, the heavy-light systems, and the light-light systems: diquarks, mesons, baryons, and tetraquarks \cite{Chen:2023djq,Chen:2023ngj,Chen:2023web}.

We notice that the general form (\ref{massfinal}) remains open and is provisional. Because there are different methods to include the masses of the light constituents. More theoretical and experimental data are needed. In addition, the parameter values are universal for both the heavy-heavy systems and the heavy-light systems \cite{Chen:2023cws,Feng:2023txx}. However, the parameter values change for the light systems \cite{Chen:2023ngj} to obtain agreeable results.

\subsection{Behaviors of the {\rts} for various baryons}\label{subsec:beh}

In this subsection, we apply the results obtained in the preceding subsection to discuss the behaviors of the {\rts} for various baryons. We find that different from the behaviors of the {\rts} for other types of baryons, the $\lambda$-trajectories and the $\rho$-trajectories for the triply heavy baryons have the same behaviors, $M{\sim}x^{2/3}_{\lambda}$ and $M{\sim}x^{2/3}_{\rho}$ or $M^2{\sim}x^{2/3}_{\lambda}$ and $M^2{\sim}x^{2/3}_{\rho}$, see Table \ref{tab:rtbehav}. Moreover, the {\rt} relations for the triply heavy baryons can be written explicitly, see Eqs. (\ref{t2q}), (\ref{pa2qQ}) and (\ref{combrt}).

As an example, we consider the baryon $(qq'){\qpp}$, in which the light diquark $(qq')$ is composed of two light quarks and ${\qpp}$ is also light.
When discussing the {\rts} of the light diquarks, Eq. (\ref{lightm3}) gives the same behavior as Eqs. (\ref{lighlight}) and (\ref{rtft}),
\bea\label{lightmrho}
M_{\rho}{\sim}x_{\rho}^{1/2},
\eea
where $M_{\rho}$ is the diquark mass.
Generally speaking, the diquark {\rts} are not the same as the $\rho$-trajectories of baryons. For a light diquark composed of two light quarks, the behavior of the diquark {\rt} is definite (see Eq. (\ref{lightmrho})) if the masses of light quarks are not considered. However, it is indefinite when discussing the $\rho$- and $\lambda$-trajectories of baryons composed of the light diquark and/or the light quark.
If the diquark $(qq')$ is regarded as being light, when Eq. (\ref{lightm3}) is chosen as the $\lambda$-trajectory relation, it will give $M{\sim}M_{\rho}^{3/2}$ while Eqs. (\ref{lighlight}) and (\ref{rtft}) gives $M{\sim}M_{\rho}$ due to different ways to include the light constituent's mass $M_{\rho}$ [corresponding to $m_{1,2}$ in Eqs. (\ref{lightm3}), (\ref{lighlight}) and (\ref{rtft}).]. They give two different behaviors of the $\rho$-trajectories. Using Eq. (\ref{lightmrho}),  Eq. (\ref{lightm3}) gives $M{\sim}x_{\rho}^{3/4}$ while Eqs. (\ref{lighlight}) and (\ref{rtft}) give $M{\sim}x_{\rho}^{1/2}$.
When discussing the $\rho$-trajectories, highly excited diquark will be involved. The first radially and orbitally excited $(ud)$ are about $1.3$ {\gev}. The masses of the $1^1p_1$ states of $(us)$ and $(ss)$ are much heavier, $1.46$ {\gev} and $1.65$ {\gev} \cite{Chen:2023ngj}, respectively. They approximate or even exceed the mass of the charm quark. Therefore, the light diquark composed of two light quarks can be regarded as being heavy when discussing the $\rho$-trajectories of baryons. Then, Eqs. (\ref{rtmeson}) and (\ref{rtft}) or Eqs. (\ref{mrtf}) and (\ref{mrfp}) can be chosen as the ${\lambda}$-trajectory formula. They will give the same $\rho$-trajectory behavior, $M{\sim}M_{\rho}$. Using Eq. (\ref{lightmrho}), we get $M{\sim}x_{\rho}^{1/2}$.

\begin{table}[!phtb]
\caption{Behaviors of two series of {\rts} for different baryons in the diquark picture. $Q=c,\,b$ and $q=u,\,d,\,s$. $M$ is the mass of the baryons. $x_{\lambda}=L,\,N_{r}$ are the angular momentum quantum number and the radial quantum number for the $\lambda$ mode. $x_{\rho}=l,\,n_{r}$ are for the $\rho$ mode. ${\bqm}$ denotes the indefinite result. }\label{tab:rtbehav}
\centering
\begin{tabular*}{0.5\textwidth}{@{\extracolsep{\fill}}cccccc@{}}
\hline\hline
    &       &   \multicolumn{2}{c}{Traj. behavior $M{\sim}x^{\nu}$} &   \multicolumn{2}{c}{Traj. behavior $M^2{\sim}x^{\nu'}$}    \\
  &    & $\rho$-mode &  $\lambda$-mode  & $\rho$-mode &  $\lambda$-mode  \\
\hline
1 & $(qq')q^{\prime\prime}$ &
$x_{\rho}^{1/2}$, $x_{\rho}^{3/4}$${\bqm}$ &
$x_{\lambda}^{1/2}$&
$x_{\rho}$, $x_{\rho}^{3/4}$${\bqm}$  &
$x_{\lambda}$,  $x_{\lambda}^{1/2}$${\bqm}$\\
2&  $(qq')Q$ &
 $x_{\rho}^{1/2}$, $x_{\rho}^{3/4}$${\bqm}$ &
 $x_{\lambda}^{1/2}$,  $x_{\lambda}^{2/3}$${\bqm}$ &
 $x_{\rho}^{1/2}$, $x_{\rho}^{3/4}$${\bqm}$ &
  $x_{\lambda}^{1/2}$,  $x_{\lambda}^{2/3}$${\bqm}$  \\
3& $(Qq)q'$ &
  $x_{\rho}^{1/2}$ &  $x_{\lambda}^{1/2}$ &
  $x_{\rho}^{1/2}$ &  $x_{\lambda}^{1/2}$ \\
4& $(Qq')Q'$ &
 $x_{\rho}^{1/2}$ & $x_{\lambda}^{2/3}$&
 $x_{\rho}^{1/2}$ &  $x_{\lambda}^{2/3}$   \\
5& $(QQ')q$     &
  $x_{\rho}^{2/3}$ &   $x_{\lambda}^{1/2}$ &
  $x_{\rho}^{2/3}$ &  $x_{\lambda}^{1/2}$\\
6& $(QQ')Q^{\prime\prime}$     &
  $x_{\rho}^{2/3}$ &  $x_{\lambda}^{2/3}$ &
  $x_{\rho}^{2/3}$ &  $x_{\lambda}^{2/3}$ \\
\hline
\hline
\end{tabular*}
\end{table}

If the light diquark is regarded as being light, $\lambda$-trajectories behave as $M{\sim}x^{1/2}_{\lambda}$ no matter if Eq. (\ref{lightm3}) or Eqs. (\ref{lighlight}) and (\ref{rtft}) are chosen as the $\lambda$-trajectory formula. If the light diquark is regarded as being heavy, $\lambda$-trajectories also behave as $M{\sim}x_{\lambda}^{1/2}$ according to Eqs. (\ref{rtmeson}) and (\ref{rtft}) or Eqs. (\ref{mrtf}) and (\ref{mrfp}). Therefore, the behavior of the $\lambda$-trajectories is definite whether the light diquark is regarded as being light or being heavy.

As in the preceding discussions, when the diquark in a baryon is composed of two light quarks, discussing the {\rts} will be confronted with two problems. One is when the light diquark can be regarded as being heavy, which will affect not only the behaviors of the ${\lambda}$-{\trs} but also those of the $\rho$-{\trs}. Another is how to introduce the mass of the light constituent if the diquark is light, which will effect the behavior of the $\rho$-{\trs}. These two problems remain open; thus, the behaviors of the {\rts} of the baryons containing the light diquark remain indefinite, see Table \ref{tab:rtbehav}.

When the diquark is doubly heavy or heavy-light, it is clear how to introduce its large mass, and the way of introducing the light mass does not affect the {\rt} behaviors (see Eqs. (\ref{massform}), (\ref{rtft}), (\ref{rtmeson}), (\ref{mrtf}) and (\ref{mrfp})). The {\rt} formulas can be written explicitly, and then the behaviors of the $\lambda$-trajectory and the $\rho$-trajectory are definite. For example, the {\rt} for the baryons composed of one doubly heavy diquark and one heavy quark can be written explicitly as Eqs. (\ref{t2q}), (\ref{pa2qQ}) and (\ref{combrt}).
Other types of baryons can be discussed in a similar manner. The {\rt} behaviors of different types of baryons are listed in Table \ref{tab:rtbehav}.

The behaviors of the {\rts} discussed are for the linearly confining potential $V_c{\sim}r$, and they vary with different confining potentials.
Moreover, the results in Table \ref{tab:rtbehav} are obtained without considering various mixings. When the mixings are considered, the behaviors of the {\rts} will become complex.

In Ref. \cite{Chen:2023web}, we suggest that all {\rts} for the diquarks, mesons, baryons, and tetraquarks are concave downwards in the $(M^2,\,x)$ $(x=L,\,N_r)$ planes. For the light systems, the {\rts} are also concave when considering the masses of the light constituents. The trajectories for baryons and tetraquarks discussed in Refs. \cite{Chen:2023web,Chen:2023djq} in fact are the $\lambda$-trajectories.
We can see from Table \ref{tab:rtbehav} that both the $\lambda$-trajectories and the $\rho$-trajectories of various baryons are all concave downwards in the $(M^2,\,x)$ $(x=l,\,n_r,\,L,\,N_r)$ planes. For the light systems, the trajectories approximate linearity as the masses are neglected and become concave when the masses are considered.

\subsection{{\rt} relations for the triply heavy baryons  }
A triply heavy baryon consists of one doubly heavy diquark $(QQ')$ and one heavy quark $Q^{\prime\prime}$. According to Eq. (\ref{massfinal}), we have the {\rt} relations for the triply heavy baryons
\begin{align}\label{t2q}
M&=m_{R{\lambda}}+\beta_{x_{\lambda}}(x_{\lambda}+c_{0x_{\lambda}})^{2/3}\;(x_{\lambda}=L,\,N_r),\nonumber\\
M_{\rho}&=m_{R\rho}+\beta_{x_{\rho}}(x_{\rho}+c_{0x_{\rho}})^{2/3}\;(x_{\rho}=l,\,n_{r}),
\end{align}
where
\begin{align}\label{pa2qQ}
m_{R{\lambda}}&=M_{\rho}+m_{Q^{\prime\prime}}+C,\nonumber\\
m_{R\rho}&=m_{Q}+m_{Q'}+C/2,\nonumber\\
\beta_{L}&=\frac{3}{2}\left(\frac{\sigma^2}{\mu_{\lambda}}\right)^{1/3}c_{fL},\; \beta_{N_r}=\frac{(3\pi)^{2/3}}{2}\left(\frac{\sigma^2}{\mu_{\lambda}}\right)^{1/3}c_{fN_r},\nonumber\\
\mu_{\lambda}&=\frac{M_{\rho}m_{Q^{\prime\prime}}}{M_{\rho}+m_{Q^{\prime\prime}}},\;
\mu_{\rho}=\frac{m_{Q}m_{Q^{\prime}}}{m_{Q}+m_{Q^{\prime}}},\nonumber\\
\beta_{l}&=\frac{3}{2}\left(\frac{\sigma^2}{4\mu_{\rho}}\right)^{1/3}c_{fl},\; \beta_{n_r}=\frac{(3\pi)^{2/3}}{2}\left(\frac{\sigma^2}{4\mu_{\rho}}\right)^{1/3}c_{fn_r}.
\end{align}
In Eq. (\ref{t2q}), $M$ is the mass of the triply heavy baryon, and $M_{\rho}$ is the mass of the doubly heavy diquark. The second relation in Eq. (\ref{t2q}) is used to calculate the diquark masses. The first relation in Eq. (\ref{t2q}) and other relations in Eqs. (\ref{t2q}) and (\ref{pa2qQ}) are employed to calculate the masses of the {\thbs}.

According to Eqs. (\ref{t2q}) and (\ref{pa2qQ}), we have
\bea
M=M_{\rho}+m_{Q^{\prime\prime}}+C+\beta_{x_{\lambda}}(x_{\lambda}+c_{0x_{\lambda}})^{2/3}
\eea
when the diquark is regarded as a constituent and the structure of it is not considered. When the diquark is considered as a bound state composed of two heavy quarks \cite{Feng:2023txx}, we have
\begin{align}\label{combrt}
M=&m_{Q}+m_{Q'}+m_{Q^{\prime\prime}}+\frac{3}{2}C\nonumber\\
&+\beta_{x_{\lambda}}(x_{\lambda}+c_{0x_{\lambda}})^{2/3}
+\beta_{x_{\rho}}(x_{\rho}+c_{0x_{\rho}})^{2/3}
\end{align}
from Eqs. (\ref{t2q}) and (\ref{pa2qQ}).
We can see from Eq. (\ref{combrt}) that there are two series of {\rts} for the triply heavy baryons: the $\lambda$-trajectories and the $\rho$-trajectories.
The {\rt} relations (Eqs. (\ref{t2q}), (\ref{pa2qQ}) and (\ref{combrt})) for the triply heavy baryons have the similar forms as the {\rt} relations for the triply heavy triquarks \cite{Song:2024bkj}.

\section{{\rts} for $\Omega_{ccb}$ and $\Omega_{cbb}$}\label{sec:rtdiquark}

In this section, both the $\lambda-${\trs} and the $\rho-${\trs} for the {\thb} are investigated.

\subsection{Parameters}
The quark masses, the string tension $\sigma$, and the parameter $C$ are from Ref. \cite{Faustov:2021qqf}. The parameters for the doubly heavy diquarks $(cc)$ and $(bb)$ are from Ref. \cite{Feng:2023txx} and listed in Table \ref{tab:parmv}. With these parameters determined, the $\rho-$modes and the diquark masses can be discussed (see Eqs. (\ref{t2q}), (\ref{pa2qQ}) and (\ref{combrt})).
To discuss the $\lambda-$modes excited states,
the parameters $c_{fL}$, $c_{fN_r}$, $c_{0{L}}$, and $c_{0N_r}$ should be determined (see Eqs. (\ref{fitcfxl}) and (\ref{fitcfxnr})). More detailed discussions are in \cite{Xie:2024dfe}.

\begin{table}[!phtb]
\caption{The values of parameters \cite{Feng:2023txx,Faustov:2021qqf}.}  \label{tab:parmv}
\centering
\begin{tabular*}{0.45\textwidth}{@{\extracolsep{\fill}}cc@{}}
\hline\hline
          & $m_{c}=1.55\; {\gev}$, \; $m_b=4.88\; {\gev}$, \\
          & $\sigma=0.18\; {\gev^2}$,\; $C=-0.3\; {\gev}$, \\
$(cc)$    & $c_{fn_{r}}=1.0$,\; $c_{fl}=1.17$,  \\
          & $c_{0n_{r}}(1^3s_1)=0.205$,\quad $c_{0{l}}(1^3s_1)=0.337$,\\
$(bb)$    & $c_{fn_{r}}=1.0$,\; $c_{fl}=1.17$,  \\
          & $c_{0n_{r}}(1^3s_1)=0.01$,\quad  $c_{0{l}}(1^3s_1)=0.001$.\\
\hline
\hline
\end{tabular*}
\end{table}

According to Eq. (\ref{t2q}), $c_{fx_{\lambda}}$ and $c_{0x_{\lambda}}$ are needed to determine a {\rt} as $m_{R_{\lambda}}$ can be calculated by using Eq. (\ref{pa2qQ}) and parameters in Table \ref{tab:parmv}. Two or more states on the {\rt} are needed to obtain $c_{fx_{\lambda}}$ and $c_{0x_{\lambda}}$. Since the experimental data are absent, we choose $c_{fx_{\lambda}}$ and $c_{0x_{\lambda}}$ by fitting these parameters from other systems. In addition, $c_{fx_{\lambda}}$ and $c_{0x_{\lambda}}$ vary with masses of constituents. We obtain the fitted parameter relations in Ref. \cite{Xie:2024dfe}
\begin{eqnarray}
c_{fL}=&1.116 + 0.013\mu_{\lambda},\; c_{0L}=0.540- 0.141\mu_{\lambda}, \label{fitcfxl}\\
c_{fN_r}=&1.008 + 0.008\mu_{\lambda}, \;  c_{0N_r}=0.334 - 0.087\mu_{\lambda},\label{fitcfxnr}
\end{eqnarray}
where $\mu_{\lambda}$ is the reduced masses, see Eq. (\ref{pa2qQ}). As more experimental data or more theoretical data become available, the fitted formulas will be refined.

\subsection{$\lambda-$ and $\rho-${\trs}}\label{subsec:rts}

\begin{table}[!phtb]
\caption{The spin-averaged masses of the $\lambda$-excited states of {\thbf} (in ${\gev}$). The notation in Eq. (\ref{tetnot}) is rewritten as $|n^{2s_d+1}l_{j_d},N^{2j+1}L_J\rangle$. And $|n^{2s_d+1}l_{j_d},NL\rangle$ denotes the spin-averaged states. Eqs. (\ref{t2q}) (or (\ref{combrt})), (\ref{fitcfxl}) and (\ref{fitcfxnr}) are used.}  \label{tab:masslambda}
\centering
\begin{tabular*}{0.50\textwidth}{@{\extracolsep{\fill}}ccc@{}}
\hline\hline
  $|n^{2s_d+1}l_{j_d},NL\rangle$        & $(cc)b$  &  $c(bb)$  \\
\hline
 $|1^3s_1, 1S\rangle$  &7.88   &11.11     \\
 $|1^3s_1, 2S\rangle$  &8.35  &11.63  \\
 $|1^3s_1, 3S\rangle$  &8.68   &11.99  \\
 $|1^3s_1, 4S\rangle$  &8.97   &12.31  \\
 $|1^3s_1, 5S\rangle$  &9.22   &12.59  \\
\hline
 $|1^3s_1, 1S\rangle$  &7.88   &11.12  \\
 $|1^3s_1, 1P\rangle$  &8.22   &11.48  \\
 $|1^3s_1, 1D\rangle$  &8.46   &11.75  \\
 $|1^3s_1, 1F\rangle$  &8.67   &11.98  \\
 $|1^3s_1, 1G\rangle$  &8.86   &12.19  \\
 $|1^3s_1, 1H\rangle$  &9.03   &12.38 \\
\hline\hline
\end{tabular*}
\end{table}

\begin{table}[!htbp]
\caption{Same as Table \ref{tab:masslambda} except for the $\rho$-excited states. $\times$ denotes the nonexist states.}  \label{tab:massrho}
\centering
\begin{tabular*}{0.5\textwidth}{@{\extracolsep{\fill}}ccc@{}}
\hline\hline
  $|n^{2s_d+1}l_{j_d},NL\rangle$        & $(cc)b$  &  $c(bb)$    \\
\hline
 $|1^3s_1, 1S\rangle$  &7.88   &11.11  \\
 $|2^3s_1, 1S\rangle$  &8.25   &11.43  \\
 $|3^3s_1, 1S\rangle$  &8.52   &11.63  \\
 $|4^3s_1, 1S\rangle$  &8.74   &11.79  \\
 $|5^3s_1, 1S\rangle$  &8.95   &11.94  \\
\hline
 $|1^3s_1, 1S\rangle$  &7.89   &11.10  \\
 $|1^3p_2, 1S\rangle$$(\times)$  &8.17   &11.36  \\
 $|1^3d_3, 1S\rangle$  &8.37   &11.51 \\
 $|1^3f_4, 1S\rangle$$(\times)$  &8.55   &11.64  \\
 $|1^3g_5, 1S\rangle$  &8.70   &11.76  \\
 $|1^3h_6, 1S\rangle$$(\times)$  &8.85   &11.86 \\
\hline\hline
\end{tabular*}
\end{table}

\begin{table*}[!phtb]
\caption{Comparison of theoretical predictions for the spin averaged masses of the {\thbs} baryons (in GeV).}  \label{tab:masscomp}
\centering
\begin{tabular*}{1.0\textwidth}{@{\extracolsep{\fill}}ccccccccc@{}}
\hline\hline
  Baryon  &  $|n^{2s_d+1}l_{j_d},NL\rangle$
          &   Our  & Ref. \cite{Faustov:2021qqf} &  Ref. \cite{Silvestre-Brac:1996myf} & Ref. \cite{Yang:2019lsg} & Ref. \cite{Roberts:2007ni} & \cite{Mathur:2018epb} &\cite{Brown:2014ena}\\
\hline
$\Omega_{ccb}$
   &$|1^3s_1, 1S\rangle$ &7.88  &7.99  &8.04  &8.02 &8.26 &8.02&8.03\\
   &$|1^3s_1, 2S\rangle$ &8.35  &8.41  &8.46  &8.46 & \\
   &$|1^3s_1, 1P\rangle$ &8.22  &8.26  &8.32  &8.31 &8.42 \\
   &$|2^3s_1, 1S\rangle$ &8.25  &8.36  &  & \\
$\Omega_{cbb}$
   &$|1^3s_1, 1S\rangle$ &11.11  &11.21  &11.24  &11.21 &11.55 &11.21&11.22\\
   &$|1^3s_1, 2S\rangle$ &11.63  &11.70  &11.64  &11.62 &\\
   &$|1^3s_1, 1P\rangle$ &11.48  &11.53  &11.55  &11.51 &11.74\\
   &$|2^3s_1, 1S\rangle$ &11.43  &11.51  &  & \\
\hline\hline
\end{tabular*}
\end{table*}

\begin{figure*}[!phtb]
\centering
\subfigure[]{\label{subfigure:cfa}\includegraphics[scale=0.48]{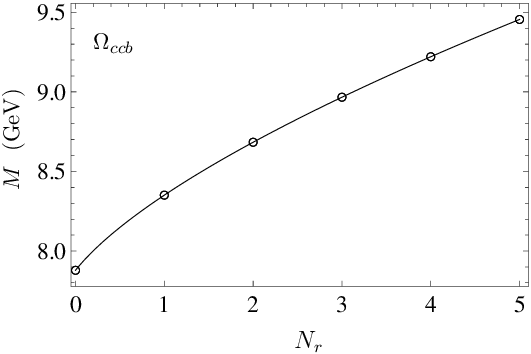}}
\subfigure[]{\label{subfigure:cfa}\includegraphics[scale=0.48]{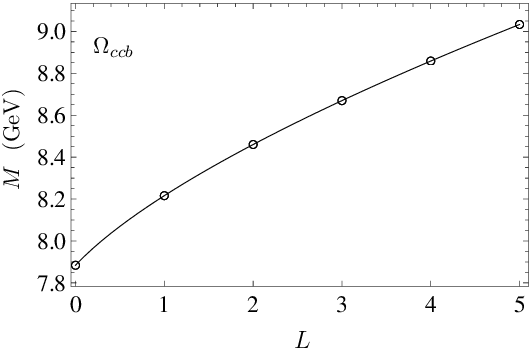}}
\subfigure[]{\label{subfigure:cfa}\includegraphics[scale=0.48]{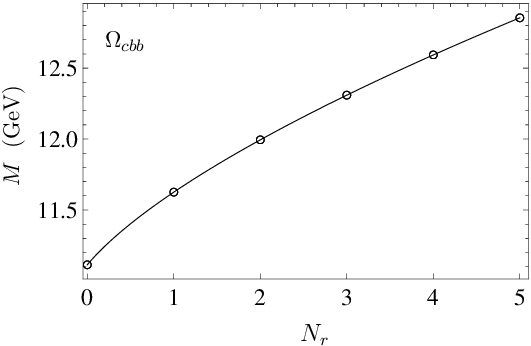}}
\subfigure[]{\label{subfigure:cfa}\includegraphics[scale=0.48]{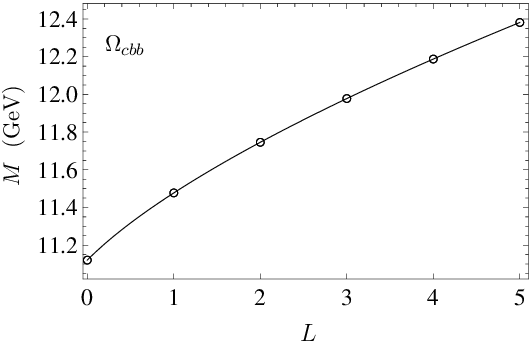}}
\caption{The $\lambda$-trajectories for the {\thbs}. Circles represent the predicted data and the black lines are the $\lambda$-trajectories, see Eq. (\ref{t2q}) or (\ref{combrt}). Data are listed in Table \ref{tab:masslambda}.}\label{fig:blam}
\end{figure*}

\begin{figure*}[!phtb]
\centering
\subfigure[]{\label{subfigure:cfa}\includegraphics[scale=0.48]{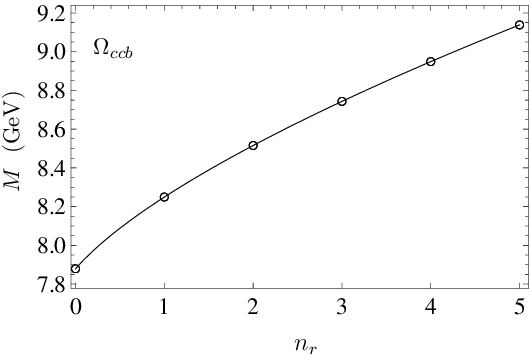}}
\subfigure[]{\label{subfigure:cfa}\includegraphics[scale=0.48]{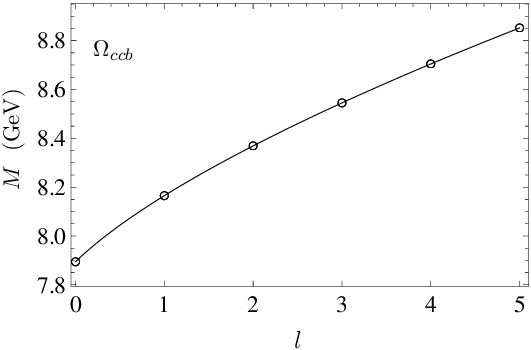}}
\subfigure[]{\label{subfigure:cfa}\includegraphics[scale=0.48]{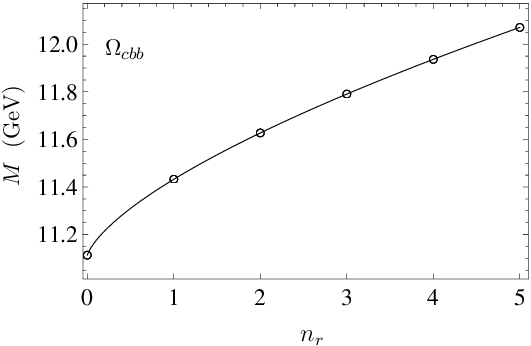}}
\subfigure[]{\label{subfigure:cfa}\includegraphics[scale=0.48]{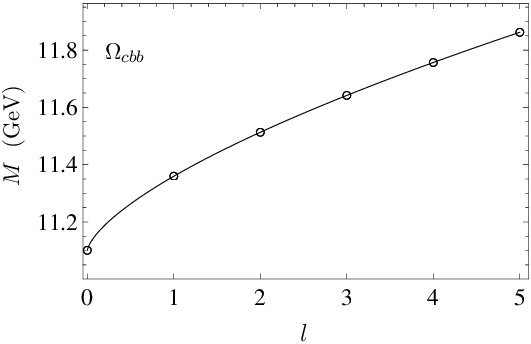}}
\caption{The $\rho$-trajectories for the {\thbs}. Circles represent the predicted data and the black lines are the $\rho$-trajectories, see Eq. (\ref{t2q}) or (\ref{combrt}). Data are listed in Table \ref{tab:massrho}.}\label{fig:brho}
\end{figure*}

By using Eqs. (\ref{t2q}) (or (\ref{combrt})), (\ref{pa2qQ}), (\ref{fitcfxl}) and (\ref{fitcfxnr}), along with the parameters in Table \ref{tab:parmv}, the spin-averaged masses of the $\lambda$-excitations and $\rho$-excitations are estimated (see Tables \ref{tab:masslambda} and \ref{tab:massrho}). The obtained results are in agreement with other theoretical predictions, see Table \ref{tab:masscomp}. The $\lambda$-{\trs} and $\rho$-{\trs} are shown in Figs. \ref{fig:blam} and \ref{fig:brho}.

There are two series of masses: one is for the $\lambda$-excited states and the other is for the $\rho$-excited states. There is no degeneracy in masses, see Tables \ref{tab:masslambda} and \ref{tab:massrho}. For an excited state, the mass of the $\lambda$-mode will be larger than that of the $\rho$-mode.
There exist mixing between $\lambda$-excitations and $\rho$-excitations or between $\lambda$-excitations themselves, for example, $3p1S-1s2P$ with $J^P=1/2^-$ and $3/2^-$ \cite{Gershtein:2000nx}. In this work, different mixings are not considered.

Both the diquarks ($(bb)$ and $(cc)$) and the quarks ($b$ and $c$) are heavy; therefore, the case of the {\thbs} is simple, and the {\rt} relations can be written explicitly, see (\ref{t2q}) and (\ref{combrt}). From Eq. (\ref{t2q}) or (\ref{combrt}), we can see easily that the $\lambda$-{\trs} and $\rho$-{\trs} have the same behaviors, $M{\sim}x^{2/3}_{\lambda}$ and $M{\sim}x^{2/3}_{\rho}$.

The curvature of the Regge trajectories holds significant
importance \cite{Chen:2018bbr}. In potential models, the curvature is related
to the dynamic equation and the confining potential.
In this work, we show that for the {\thb}, not only the $\lambda$-trajectories but also the $\rho$-trajectories are concave in the $(M^2,\,x)$ planes, see Eq. (\ref{t2q}) or (\ref{combrt}).

\section{Conclusions}\label{sec:conc}

By employing the diquark {\rt} relation, we present explicitly the {\rt} relations for the {\thbs}, which can investigate both the $\lambda$-mode excited states and the $\rho$-mode excited states. We estimate the masses of the $\lambda$-excited states and the $\rho$-excited states. The results are in agreement with other theoretical predictions.

Both the $\lambda$-trajectories and the $\rho$-trajectories are discussed.
The $\rho$-trajectories behave as $M{\sim}x_{\rho}^{2/3}$ because the $\rho$-mode excitations are in the doubly heavy diquarks ($(bb)$ or $(cc)$), which are the heavy-heavy systems. Similarly, the $\lambda$-trajectories behave as $M{\sim}x_{\lambda}^{2/3}$ because the $\lambda$-mode excitations occur between the doubly heavy diquark and the heavy quark, with the diquark ($(cc)$ or $(bb)$) and quark ($b$ or $c$) forming a heavy-heavy system.

Moreover, the behaviors of the $\lambda$- and $\rho$-trajectories for various baryons are discussed.
It is shown that both the $\lambda$-trajectories and the $\rho$-trajectories for baryons are concave downwards in the $(M^2,\,x)$ plane. The Regge trajectories for the light baryons approximate linearity and become concave as the masses of the light constituents are considered.

\section*{Acknowledgments}
We are very grateful to the anonymous referees for the valuable comments and suggestions.

\vspace{5mm}
\noindent{\bf Data Availability Statement} This manuscript has no associated data. [Author's comment: All data are included in the manuscript.].

\vspace{5mm}
\noindent{\bf Code Availability Statement} This manuscript has no associated code/software. [Author's comment: The manuscript has no associated code/software.].

\vspace{5mm}
\noindent{\bf Open Access} This article is licensed under a Creative Commons Attribution 4.0 International License, which permits use, sharing, adaptation, distribution and reproduction in any medium or format, as long as you give appropriate credit to the original author(s) and the source, provide a link to the Creative Commons licence, and indicate if changes were made. The images or other third party material in this article are included in the article's Creative Commons licence, unless indicated otherwise in a credit line to the material. If material is not included in the article's Creative Commons licence and your intended use is not permitted by statutory regulation or exceeds the permitted use, you will need to obtain permission directly from the copyright holder. To view a copy of this licence, visit \text{http://creativecommons.org/licenses/by/4.0/}.

\noindent{Funded by SCOAP$^3$.}

\appendix

\section{States of baryons}
The states of baryons in the diquark picture are listed in Table \ref{tab:bqstates}.

\begin{table*}[!phtb]
\caption{The states of baryons in the diquark picture. The notation is explained in \ref{subsec:prelim}. Here, $q$, $\qp$ and $\qpp$ represent both the light quarks and the heavy quarks.}  \label{tab:bqstates}
\centering
\begin{tabular*}{\textwidth}{@{\extracolsep{\fill}}ccccc@{}}
\hline\hline
$J^P$ & $(L,l)$  &  Configuration \\
\hline
$\frac{1}{2}^+$ & $(0,0)$ & $\left({\qqs}_{n^1s_0}{\qpp}\right)_{N^2S_{1/2}}$,\;
$\left({\qqb}_{n^3s_1}{\qpp}\right)_{N^2S_{1/2}}$,\\
& $(1,1)$ &
$\left({\qqb}_{n^1p_1}{\qpp}\right)_{N^2P_{1/2}}$,\;
$\left({\qqb}_{n^1p_1}{\qpp}\right)_{N^4P_{1/2}}$,\;
$\left({\qqs}_{n^3p_0}{\qpp}\right)_{N^2P_{1/2}}$,\;
$\left({\qqs}_{n^3p_1}{\qpp}\right)_{N^2P_{1/2}}$,\\
& &
$\left({\qqs}_{n^3p_1}{\qpp}\right)_{N^4P_{1/2}}$,\;
$\left({\qqs}_{n^3p_2}{\qpp}\right)_{N^4P_{1/2}}$\\
&$\cdots$ &$\cdots$ \\
$\frac{1}{2}^-$ & $(1,0)$ &
$\left({\qqs}_{n^1s_0}{\qpp}\right)_{N^2P_{1/2}}$,\;
$\left({\qqb}_{n^3s_1}{\qpp}\right)_{N^2P_{1/2}}$,\;
$\left({\qqb}_{n^3s_1}{\qpp}\right)_{N^4P_{1/2}}$,\\
& $(0,1)$ &
$\left({\qqb}_{n^1p_1}{\qpp}\right)_{N^2S_{1/2}}$,\;
$\left({\qqs}_{n^3p_0}{\qpp}\right)_{N^2S_{1/2}}$,\;
$\left({\qqs}_{n^3p_1}{\qpp}\right)_{N^2S_{1/2}}$,\\
&$\cdots$ &$\cdots$ \\
$\frac{3}{2}^+$ & $(0,0)$ &
$\left({\qqb}_{n^3s_1}{\qpp}\right)_{N^4S_{3/2}}$,\\
& $(1,1)$ &
$\left({\qqb}_{n^1p_1}{\qpp}\right)_{N^2P_{3/2}}$,\;
$\left({\qqb}_{n^1p_1}{\qpp}\right)_{N^4P_{3/2}}$,\;
$\left({\qqs}_{n^3p_0}{\qpp}\right)_{N^2P_{3/2}}$,\;
$\left({\qqs}_{n^3p_1}{\qpp}\right)_{N^2P_{3/2}}$,\;\\
& &
$\left({\qqs}_{n^3p_1}{\qpp}\right)_{N^4P_{3/2}}$,\;
$\left({\qqs}_{n^3p_2}{\qpp}\right)_{N^4P_{3/2}}$,\;
$\left({\qqs}_{n^3p_2}{\qpp}\right)_{N^6P_{3/2}}$,\\
&$\cdots$ &$\cdots$ \\
$\frac{3}{2}^-$ & $(1,0)$ &
$\left({\qqs}_{n^1s_0}{\qpp}\right)_{N^2P_{3/2}}$,\;
$\left({\qqb}_{n^3s_1}{\qpp}\right)_{N^2P_{3/2}}$,\;
$\left({\qqb}_{n^3s_1}{\qpp}\right)_{N^4P_{3/2}}$,\\
& $(0,1)$ &
$\left({\qqb}_{n^1p_1}{\qpp}\right)_{N^4S_{3/2}}$,\;
$\left({\qqs}_{n^3p_1}{\qpp}\right)_{N^4S_{3/2}}$,\;
$\left({\qqs}_{n^3p_2}{\qpp}\right)_{N^4S_{3/2}}$\\
&$\cdots$ &$\cdots$ \\
\hline\hline
\end{tabular*}
\end{table*}


\begin{thebibliography}{99}


\bibitem{Mathur:2018epb}
N.~Mathur, M.~Padmanath and S.~Mondal,
Phys. Rev. Lett. \textbf{121}, no.20, 202002 (2018)
doi:10.1103/PhysRevLett.121.202002
[arXiv:1806.04151 [hep-lat]].

\bibitem{Brown:2014ena}
Z.~S.~Brown, W.~Detmold, S.~Meinel and K.~Orginos,
Phys. Rev. D \textbf{90}, no.9, 094507 (2014)
doi:10.1103/PhysRevD.90.094507
[arXiv:1409.0497 [hep-lat]].

\bibitem{Serafin:2018aih}
K.~Serafin, M.~G\'omez-Rocha, J.~More and S.~D.~G\l{}azek,
Eur. Phys. J. C \textbf{78}, no.11, 964 (2018)
doi:10.1140/epjc/s10052-018-6436-2
[arXiv:1805.03436 [hep-ph]].


\bibitem{Faustov:2021qqf}
R.~N.~Faustov and V.~O.~Galkin,
Phys. Rev. D \textbf{105}, no.1, 014013 (2022)
doi:10.1103/PhysRevD.105.014013
[arXiv:2111.07702 [hep-ph]].


\bibitem{Silvestre-Brac:1996myf}
B.~Silvestre-Brac,
Few Body Syst. \textbf{20}, 1-25 (1996)
doi:10.1007/s006010050028



\bibitem{Roberts:2007ni}
W.~Roberts and M.~Pervin,
Int. J. Mod. Phys. A \textbf{23}, 2817-2860 (2008)
doi:10.1142/S0217751X08041219
[arXiv:0711.2492 [nucl-th]].


\bibitem{deArenaza:2024dhe}
N.~M.~de Arenaza, J.~J.~G\'alvez-Viruet and F.~J.~Llanes-Estrada,
[arXiv:2407.07232 [nucl-th]].


\bibitem{Yang:2019lsg}
G.~Yang, J.~Ping, P.~G.~Ortega and J.~Segovia,
Chin. Phys. C \textbf{44}, no.2, 023102 (2020)
doi:10.1088/1674-1137/44/2/023102
[arXiv:1904.10166 [hep-ph]].

\bibitem{Hasenfratz:1980ka}
P.~Hasenfratz, R.~R.~Horgan, J.~Kuti and J.~M.~Richard,
Phys. Lett. B \textbf{94}, 401-404 (1980)
doi:10.1016/0370-2693(80)90906-5


\bibitem{Zhang:2009re}
J.~R.~Zhang and M.~Q.~Huang,
Phys. Lett. B \textbf{674}, 28-35 (2009)
doi:10.1016/j.physletb.2009.02.056
[arXiv:0902.3297 [hep-ph]].


\bibitem{Najjar:2024deh}
Z.~R.~Najjar, K.~Azizi and H.~R.~Moshfegh,
Eur. Phys. J. C \textbf{84}, no.6, 612 (2024)
doi:10.1140/epjc/s10052-024-12960-x
[arXiv:2402.14348 [hep-ph]].

\bibitem{Wang:2011ae}
Z.~G.~Wang,
Commun. Theor. Phys. \textbf{58}, 723-731 (2012)
doi:10.1088/0253-6102/58/5/17
[arXiv:1112.2274 [hep-ph]].


\bibitem{Martynenko:2007je}
A.~P.~Martynenko,
Phys. Lett. B \textbf{663}, 317-321 (2008)
doi:10.1016/j.physletb.2008.04.030
[arXiv:0708.2033 [hep-ph]].


\bibitem{Jia:2006gw}
Y.~Jia,
JHEP \textbf{10}, 073 (2006)
doi:10.1088/1126-6708/2006/10/073
[arXiv:hep-ph/0607290 [hep-ph]].

\bibitem{Yin:2019bxe}
P.~L.~Yin, C.~Chen, G.~Krein, C.~D.~Roberts, J.~Segovia and S.~S.~Xu,
Phys. Rev. D \textbf{100}, no.3, 034008 (2019)
doi:10.1103/PhysRevD.100.034008
[arXiv:1903.00160 [nucl-th]].


\bibitem{Ishida:1994pf}
S.~Ishida, M.~Ishida and K.~Yamada,
Prog. Theor. Phys. \textbf{91}, 775-800 (1994)
doi:10.1143/PTP.91.775


\bibitem{Chen:2023djq}
J.~K.~Chen,
Nucl. Phys. A \textbf{1050}, 122927 (2024)
doi:10.1016/j.nuclphysa.2024.122927
[arXiv:2302.05926 [hep-ph]].


\bibitem{Oudichhya:2023pkg}
J.~Oudichhya, K.~Gandhi and A.~k.~Rai,
Pramana \textbf{97}, no.4, 151 (2023)
doi:10.1007/s12043-023-02630-0
[arXiv:2304.05110 [hep-ph]].


\bibitem{Chen:2022flh}
J.~K.~Chen,
Nucl. Phys. B \textbf{983}, 115911 (2022)
doi:10.1016/j.nuclphysb.2022.115911
[arXiv:2203.02981 [hep-ph]].

\bibitem{Chen:2023cws}
J.~K.~Chen, X.~Feng and J.~Q.~Xie,
JHEP \textbf{10}, 052 (2023)
doi:10.1007/JHEP10(2023)052
[arXiv:2308.02289 [hep-ph]].


\bibitem{Burns:2010qq}
T.~J.~Burns, F.~Piccinini, A.~D.~Polosa and C.~Sabelli,
Phys. Rev. D \textbf{82}, 074003 (2010)
doi:10.1103/PhysRevD.82.074003
[arXiv:1008.0018 [hep-ph]].



\bibitem{Chen:2021kfw}
J.~K.~Chen,
Eur. Phys. J. A \textbf{57}, 238 (2021)
doi:10.1140/epja/s10050-021-00502-y
[arXiv:2102.07993 [hep-ph]].

\bibitem{Chen:2018nnr}
J.~K.~Chen,
Eur. Phys. J. C \textbf{78}, no.8, 648 (2018)
doi:10.1140/epjc/s10052-018-6134-0


\bibitem{Chen:2023web}
J.~K.~Chen,
Eur. Phys. J. C \textbf{84}, no.4, 356 (2024)
doi:10.1140/epjc/s10052-024-12706-9
[arXiv:2302.06794 [hep-ph]].


\bibitem{Xie:2024dfe}
J.~Q.~Xie, H.~Song, X.~Feng and J.~K.~Chen,
[arXiv:2407.04222 [hep-ph]].



\bibitem{Feng:2023txx}
X.~Feng, J.~K.~Chen and J.~Q.~Xie,
Phys. Rev. D \textbf{108}, no.3, 034022 (2023)
doi:10.1103/PhysRevD.108.034022
[arXiv:2305.15705 [hep-ph]].




\bibitem{Chen:2023ngj}
J.~K.~Chen, J.~Q.~Xie, X.~Feng and H.~Song,
Eur. Phys. J. C \textbf{83}, no.12, 1133 (2023)
doi:10.1140/epjc/s10052-023-12329-6
[arXiv:2310.05131 [hep-ph]].





\bibitem{Bedolla:2019zwg}
M.~A.~Bedolla, J.~Ferretti, C.~D.~Roberts and E.~Santopinto,
Eur. Phys. J. C \textbf{80}, no.11, 1004 (2020)
doi:10.1140/epjc/s10052-020-08579-3
[arXiv:1911.00960 [hep-ph]].

\bibitem{Ferretti:2019zyh}
J.~Ferretti,
Few Body Syst. \textbf{60}, no.1, 17 (2019)
doi:10.1007/s00601-019-1483-2

\bibitem{Godfrey:1985xj}
S.~Godfrey and N.~Isgur,
Phys. Rev. D \textbf{32}, 189-231 (1985)
doi:10.1103/PhysRevD.32.189



\bibitem{Durand:1981my}
B.~Durand and L.~Durand,
Phys. Rev. D \textbf{25}, 2312 (1982)
doi:10.1103/PhysRevD.25.2312

\bibitem{Durand:1983bg}
B.~Durand and L.~Durand,
Phys. Rev. D \textbf{30}, 1904 (1984)
doi:10.1103/PhysRevD.30.1904

\bibitem{Lichtenberg:1982jp}
D.~B.~Lichtenberg, W.~Namgung, E.~Predazzi and J.~G.~Wills,
Phys. Rev. Lett. \textbf{48}, 1653 (1982)
doi:10.1103/PhysRevLett.48.1653


\bibitem{Jacobs:1986gv}
S.~Jacobs, M.~G.~Olsson and C.~Suchyta, III,
Phys. Rev. D \textbf{33}, 3338 (1986)
[erratum: Phys. Rev. D \textbf{34}, 3536 (1986)]
doi:10.1103/PhysRevD.33.3338


\bibitem{Faustov:2021hjs}
R.~N.~Faustov, V.~O.~Galkin and E.~M.~Savchenko,
Universe \textbf{7}, no.4, 94 (2021)
doi:10.3390/universe7040094
[arXiv:2103.01763 [hep-ph]].




\bibitem{Lundhammar:2020xvw}
P.~Lundhammar and T.~Ohlsson,
Phys. Rev. D \textbf{102}, no.5, 054018 (2020)
doi:10.1103/PhysRevD.102.054018
[arXiv:2006.09393 [hep-ph]].



\bibitem{Ferretti:2011zz}
J.~Ferretti, A.~Vassallo and E.~Santopinto,
Phys. Rev. C \textbf{83}, 065204 (2011)
doi:10.1103/PhysRevC.83.065204

\bibitem{Eichten:1974af}
E.~Eichten, K.~Gottfried, T.~Kinoshita, J.~B.~Kogut, K.~D.~Lane and T.~M.~Yan,
Phys. Rev. Lett. \textbf{34}, 369-372 (1975)
[erratum: Phys. Rev. Lett. \textbf{36}, 1276 (1976)]
doi:10.1103/PhysRevLett.34.369


\bibitem{Lucha:1991vn}
W.~Lucha, F.~F.~Schoberl and D.~Gromes,
Phys. Rept. \textbf{200}, 127-240 (1991)
doi:10.1016/0370-1573(91)90001-3


\bibitem{Gromes:1981cb}
D.~Gromes,
Z. Phys. C \textbf{11}, 147 (1981)
doi:10.1007/BF01573997



\bibitem{Brau:2000st}
F.~Brau,
Phys. Rev. D \textbf{62}, 014005 (2000)
doi:10.1103/PhysRevD.62.014005
[arXiv:hep-ph/0412170 [hep-ph]].

\bibitem{Veseli:1996gy}
S.~Veseli and M.~G.~Olsson,
Phys. Lett. B \textbf{383}, 109-115 (1996)
doi:10.1016/0370-2693(96)00721-6
[arXiv:hep-ph/9606257 [hep-ph]].


\bibitem{Selem:2006nd}
A.~Selem and F.~Wilczek,
doi:10.1142/9789812773524\_0030
[arXiv:hep-ph/0602128 [hep-ph]].


\bibitem{Nielsen:2018uyn}
M.~Nielsen and S.~J.~Brodsky,
Phys. Rev. D \textbf{97}, no.11, 114001 (2018)
doi:10.1103/PhysRevD.97.114001
[arXiv:1802.09652 [hep-ph]].

\bibitem{Chen:2014nyo}
B.~Chen, K.~W.~Wei and A.~Zhang,
Eur. Phys. J. A \textbf{51}, 82 (2015)
doi:10.1140/epja/i2015-15082-3
[arXiv:1406.6561 [hep-ph]].


\bibitem{Sonnenschein:2018fph}
J.~Sonnenschein and D.~Weissman,
Eur. Phys. J. C \textbf{79}, no.4, 326 (2019)
doi:10.1140/epjc/s10052-019-6828-y
[arXiv:1812.01619 [hep-ph]].



\bibitem{MartinContreras:2020cyg}
M.~A.~Martin Contreras and A.~Vega,
Phys. Rev. D \textbf{102}, no.4, 046007 (2020)
doi:10.1103/PhysRevD.102.046007
[arXiv:2004.10286 [hep-ph]].


\bibitem{Afonin:2014nya}
S.~S.~Afonin and I.~V.~Pusenkov,
Phys. Rev. D \textbf{90}, no.9, 094020 (2014)
doi:10.1103/PhysRevD.90.094020
[arXiv:1411.2390 [hep-ph]].


\bibitem{Sergeenko:1994ck}
M.~N.~Sergeenko,
Z. Phys. C \textbf{64}, 315-322 (1994)
doi:10.1007/BF01557404


\bibitem{Gershtein:2000nx}
S.~S.~Gershtein, V.~V.~Kiselev, A.~K.~Likhoded and A.~I.~Onishchenko,
Phys. Rev. D \textbf{62}, 054021 (2000)
doi:10.1103/PhysRevD.62.054021

\bibitem{Song:2024bkj}
H.~Song, J.~Q.~Xie and J.~K.~Chen,
[arXiv:2408.03720 [hep-ph]].

\bibitem{Chen:2018bbr}
J.~K.~Chen,
Phys. Lett. B \textbf{786}, 477-484 (2018)
doi:10.1016/j.physletb.2018.10.022
[arXiv:1807.11003 [hep-ph]].


\end{thebibliography}
\end{document}